\documentclass[conference]{IEEEtran}
\IEEEoverridecommandlockouts

\usepackage[hyphens]{url}

\usepackage{hyperref}

\usepackage{xspace}

\usepackage{amsfonts}

\usepackage{cite}
\usepackage{amsmath,amssymb,amsfonts}
\usepackage{algorithmic}
\usepackage{graphicx}
\usepackage{textcomp}
\usepackage{xcolor}
\def\BibTeX{{\rm B\kern-.05em{\sc i\kern-.025em b}\kern-.08em
    T\kern-.1667em\lower.7ex\hbox{E}\kern-.125emX}}

\makeatletter 
\newcommand{\linebreakand}{%
  \end{@IEEEauthorhalign}
  \hfill\mbox{}\par
  \mbox{}\hfill\begin{@IEEEauthorhalign}
}
\makeatother 

\usepackage{flushend}
    
\begin{document}

\title{Neural network accelerator for quantum control}

\author{
\IEEEauthorblockN{David Xu}
\IEEEauthorblockA{\textit{Columbia University}\\
New York, USA\\
dx2199@columbia.edu}
\and
\IEEEauthorblockN{A. Bar\i\c{s} \"Ozg\"uler}
\IEEEauthorblockA{\textit{Fermilab} \\
Batavia, USA\\
aozguler@fnal.gov}
\and
\IEEEauthorblockN{Giuseppe Di Guglielmo}
\IEEEauthorblockA{\textit{Fermilab} \\
Batavia, USA\\
gdg@fnal.gov}
\and
\linebreakand
\IEEEauthorblockN{Nhan Tran}
\IEEEauthorblockA{\textit{Fermilab} \\
Batavia, USA\\
ntran@fnal.gov}
\and
\IEEEauthorblockN{Gabriel N. Perdue}
\IEEEauthorblockA{\textit{Fermilab} \\
Batavia, USA\\
perdue@fnal.gov}
\and
\IEEEauthorblockN{Luca Carloni}
\IEEEauthorblockA{\textit{Columbia University}\\
New York, USA\\
luca@cs.columbia.edu}
\and
\IEEEauthorblockN{Farah Fahim}
\IEEEauthorblockA{\textit{Fermilab} \\
Batavia, USA\\
farah@fnal.gov}
\thanks{Corresponding author: A. Bar\i\c{s} \"Ozg\"uler, aozguler@fnal.gov}
}

\maketitle

\thispagestyle{plain}
\pagestyle{plain}

\begin{abstract}
Efficient quantum control is necessary for practical quantum computing implementations with current technologies. Conventional algorithms for determining optimal control parameters are computationally expensive, largely excluding them from use outside of the simulation. Existing hardware solutions structured as lookup tables are imprecise and costly. By designing a machine learning model to approximate the results of traditional tools, a more efficient method  can be produced. Such a model can then be synthesized into a hardware accelerator for use in quantum systems. In this study, we demonstrate a machine learning algorithm for predicting optimal pulse parameters. This algorithm is lightweight enough to fit on a low-resource FPGA and perform inference with a latency of 175\,ns and pipeline interval of 5\,ns with $~>~$0.99 gate fidelity. In the long term, such an accelerator could be used near quantum computing hardware where traditional computers cannot operate, enabling quantum control at a reasonable cost at low latencies without incurring large data bandwidths outside of the cryogenic environment. 
\end{abstract}


\section{Introduction}
\label{sec:intro}

Quantum computing has great potential but today's quantum computers are still encumbered by errors that limit their practical usefulness~\cite{fellous20211, ball20211}. Quantum control seeks to enable  precise manipulation of quantum hardware via control-theoretical approaches~\cite{ozguler2022numerical, koch2022quantum, ozguler2022dynamics}. The result would be an increase in performance of the imperfect hardware. It is desirable to apply such techniques in actual quantum-computational environments, necessitating purpose-built control hardware that can function at low temperature~\cite{butko20201, xue2021cmos, koolstra2022monitoring, stefanazzi2022qick, alam2022quantum}. By approximating existing quantum control algorithms using machine learning (ML), we hope to achieve comparable functionality using significantly less computation. The resulting model could then be implemented as a hardware accelerator at reasonable costs~\cite{overwater2022neural}.

A qubit (quantum bit) is a quantum computing system's smallest unit of information. Its quantum state can be described as a linear combination of two orthonormal states $\left|0\right\rangle$ and $\left|1\right\rangle$: $A$ = $x$ $\left|0\right\rangle$ + $y$ $\left|1\right\rangle$, where $x$ and $y$ are complex numbers. Quantum gates apply specified transformations, changing  quantum states accordingly. We desire to create a target gate $U$ which optimally changes quantum state $A$ into $U*A$. Time-varying pulses are used to produce such action. For some gate, $U(\beta)$, where $\beta$ is a gate parameter, the problem of gate synthesis can be understood as finding optimal pulse parameters $\alpha$ to produce action $U(\beta)$. Devising pulses for a given value of $\beta$ is computationally expensive because the size of the underlying pulse parameters (the vector $\alpha$) can be quite large. Furthermore, there are infinitely many possible $\beta$ (as generally $\beta$ is selected from some subset of $\mathbb{R}$). Conventional tools for such calculations include Juqbox.jl, Quandary, and QuTiP~\cite{juqboxjl, quandary, qutip}. These tools utilize optimization techniques such as gradient ascent to discover optimal $\alpha$.

\begin{figure}[t]
\centering
\includegraphics[width=0.95\columnwidth]{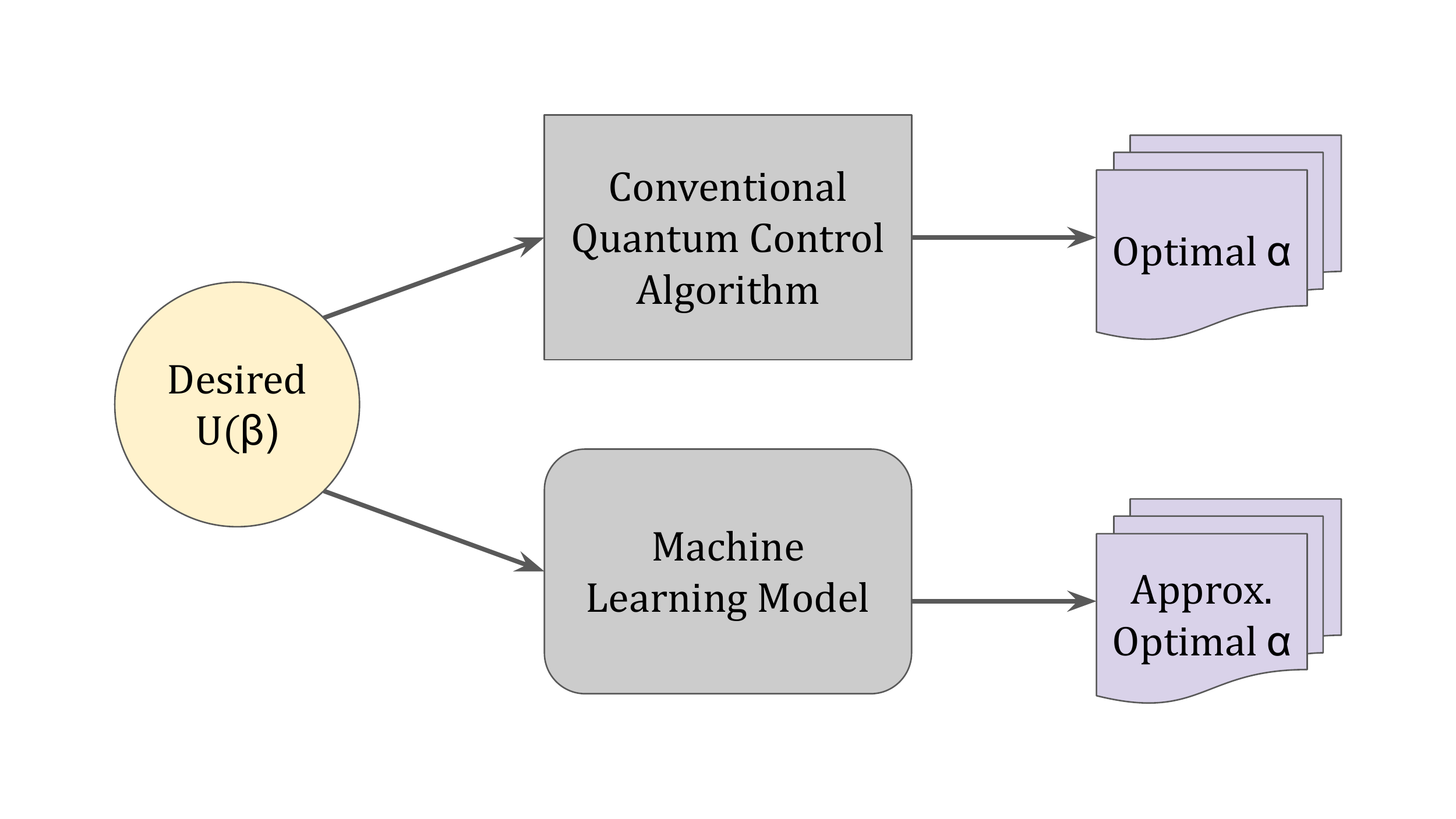}
\caption{Simplified overview.}
\label{fig:overview}
\end{figure}

In this work, a machine learning model will be trained to determine the appropriate $\alpha$ for producing $U(\beta)$ for any $\beta$ (Fig.~\ref{fig:overview}). The model will be implemented as a hardware accelerator via {\tt{hls4ml}}~\cite{hls4ml}. hls4ml is a toolset that converts pre-trained ML models into accelerators for FPGA or ASIC implementation with low power or ultra-low latency goals. It requires little designer overhead. Furthermore, the integration with the Google QKeras library allows aggressively quantized deep neural networks to be implemented. The final model is highly resource-efficient and sacrifices little to no accuracy. 
Since the goal of our present work is to produce an accelerator capable of operating within the harsh conditions of a quantum computing environment, it is desirable for such an accelerator to be small to minimize energy dissipation within the low-temperature quantum environments managed by systems of limited cooling capacity. Hardware implementations  need to be fast so that quantum computations can be done within the coherence time since noise is a big bottleneck for current quantum devices.

Note that hardware implementations that function as lookup tables mapping from a fixed number of $\beta$ to the corresponding $\alpha$ do exist \cite{liang20221}. However, such tables are limited by their precision, as only entries for a finite number of $\beta$ can be stored. Further discussion of the advantages and disadvantages of such systems can be found in Section~\ref{sec:lut}.

\begin{figure}[t]
\centering
\includegraphics[width=0.95\columnwidth]{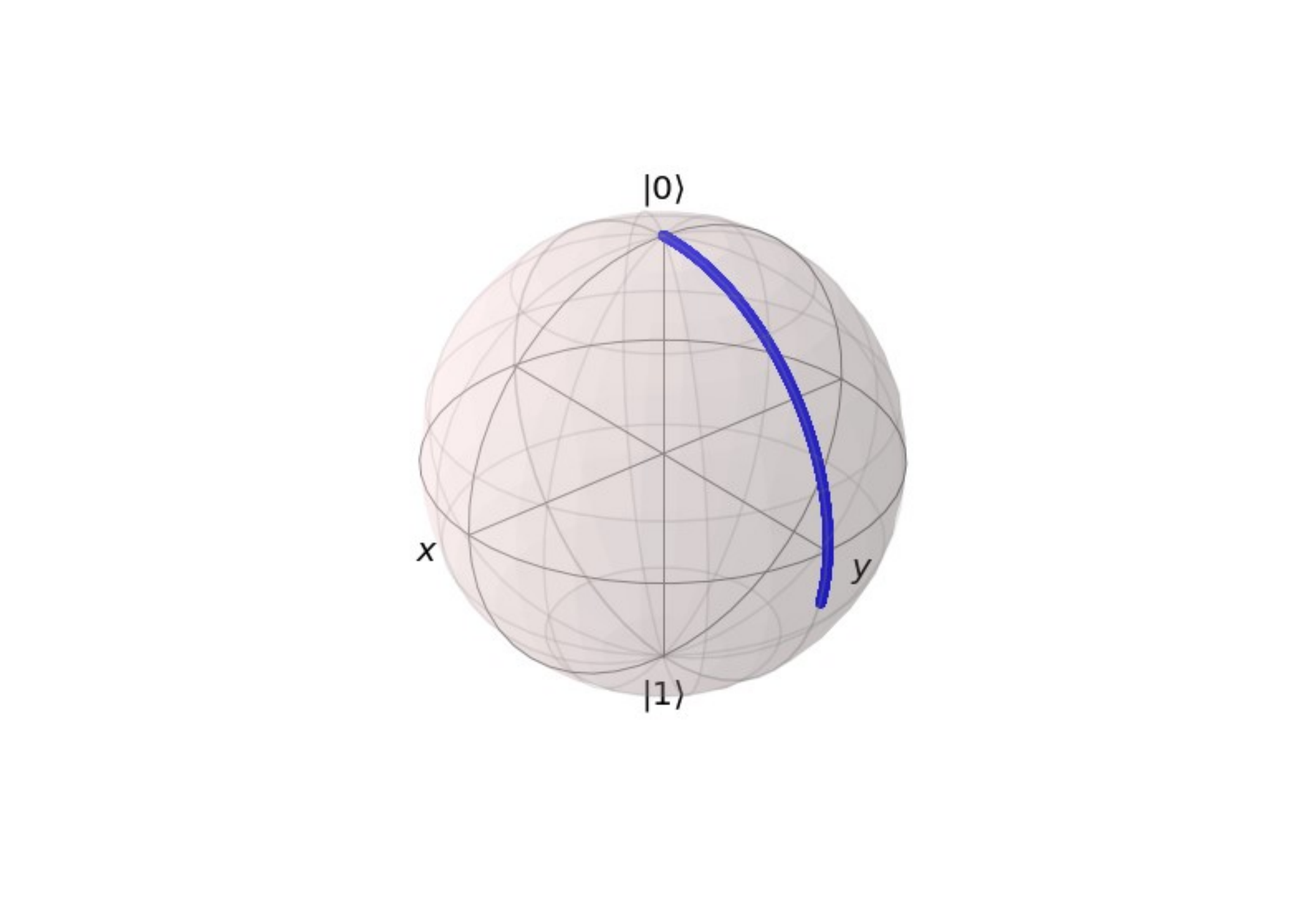}
\caption{Bloch sphere visualization of a rotation transformation starting from the $\left|0\right\rangle$ quantum state.}
\label{fig:transformsample}
\end{figure}

For this exploration, a simplified scenario is specified as follows: The gate $U(\beta)$ (X-gate) rotates a qubit about the x-axis by angle $-\pi{}\leq{}\beta{}\leq{}\pi{}$ with pulse duration 100 ns. We model the qubit as the ground and first excited states of a transmon, a type of superconducting qubit. The pulses are specified by a 20-dimensional vector $\alpha$. The dimension ``20'' is due to the B-spline parametrization we follow in Juqbox \cite{juqboxjl}, which is based on the numerical techniques in Refs. \cite{petersson2020discrete, petersson2021optimal}. The dimension of the vector $\alpha$ is proportional to the multiplication of number of carrier frequencies and number of B-splines (here, 1 and 10, respectively). X-gate brings the state (ground state) to a superposition state as shown in Figure~\ref{fig:transformsample}, which visualizes on a Bloch sphere the qubit being rotated from its initial $\left|0\right\rangle$ quantum state.

The accelerator will be evaluated for accuracy, cost, and performance. Accuracy is measured by comparing the ML-model results to those of conventional quantum control algorithms. This metric is considered both for the software model and its synthesized hardware implementation, which may perform differently due to quantization. Accuracy evaluation includes both traditional ML measures such as mean squared error (MSE) and domain-specific measures such as quantum gate fidelity. Cost is measured as the FPGA area of the synthesized result. Performance is measured via an analysis of accelerator latency and throughput.

Fidelity is a metric of the overlap between two unitaries (gates), $U_1$ and $U_2$. It can be defined in several ways. We use the definition \cite{pedersen20071}:
\begin{equation}
F= \displaystyle \frac{M+\left|tr({U})\right|^2}{M(M+1)}\label{eq:QuantumFidelity},
\end{equation} 
where $U = U_1^\dagger U_2$ and $M=2$ is the dimensionality of the Hilbert space for single qubit. We use the following definitions for three types of unitaries considered in this paper. The ``Golden'' gate is the target gate. "Optimized" gate is calculated via pulse optimizer (\texttt{Juqbox.jl} \cite{juqboxjl} for this paper), which starts from a random pulse and finds the optimal pulse until reaching a target fidelity (calculated as overlap between "optimized" and "golden" gates). The ``Predicted'' gate is the one calculated from the ML model. We check the overlap between these three types of gates, where $U_1$ and $U_2$ are two of these three types (golden, optimized, predicted).

\section{Model development}
\label{sec:model_development}

Our initial dataset consists of 101 samples. Each sample is a rotation angle upon the x-axis of a qubit ($\beta$ value) and the corresponding quantum state transformations in the form of 20 parameters ($\alpha$ vector) that were generated with Juqbox, i.e., an open-source software package designed to solve quantum optimal control problems in closed systems~\cite{juqboxjl}. Our later analysis has shown that the models perform well for most angles other than $-\pi$. This was because the dataset used the same pulses for the angles $-\pi$ and $\pi$, as both transformations produce the same final quantum state (i.e., the transformation for $-\pi$ actually is carried out in the positive direction). Thus we decided to remove from the dataset the samples that were associated with $-\pi$. We used a 60-20-20-percent split for training, validation, and test sets.

\begin{figure}[t!]
\centering
\includegraphics[width=0.95\columnwidth]{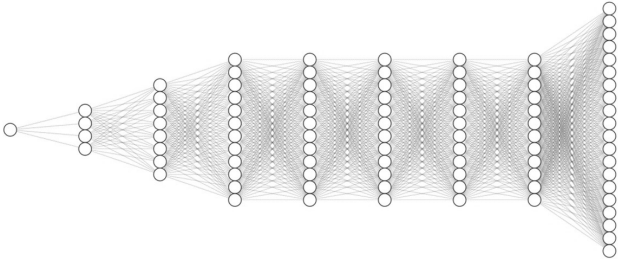}
\caption{The initial model architecture: a multi-layer perceptron with seven hidden layers, 1,040 parameters.\label{fig:mlp_forward}}
\end{figure}

As a starting point in our neural-architecture exploration, we chose the multi-layer perceptron (MLP) shown in Fig.~\ref{fig:mlp_forward}. The model has a total of seven hidden layers and 1,040 trainable parameters. We trained the model in Tensorflow with Adam optimizer and minimized the mean-squared error (MSE) as a loss function. We set a target of $5,000$ epochs with an early-stopping callback~\cite{earlystopping}.
The final MSE value that we measured on the test set was $7.908\times{}10^{-8}$. Since there is no ``correct'' value for MSE, but the lower, the better, and a zero value means the model is perfect, we assumed this value as our reference for the rest of the neural-architecture exploration.

\begin{figure}[t]
\centering
\includegraphics[width=0.95\columnwidth]{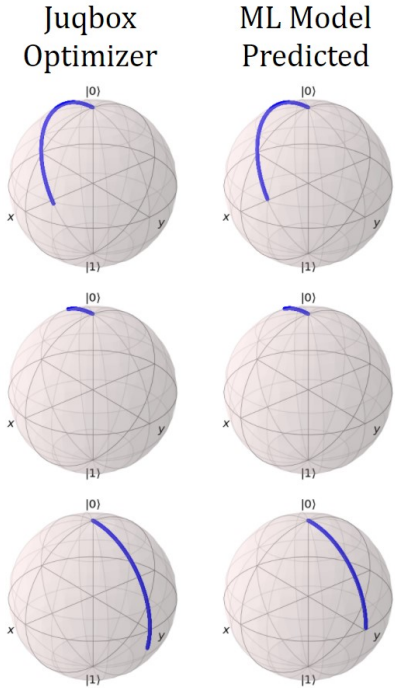}
\caption{Comparison of transformations produced by Juqbox optimizer and our machine-learning model. Each row illustrates results for a different angle $\beta{}$.}
\label{fig:transformcomparison}
\end{figure}

We also compared the expected and predicted results of the pulses on the Bloch sphere using the graphical features of the QuTiP framework~\cite{qutip}. Fig.~\ref{fig:transformcomparison} shows the similarities of the results. This visual comparison further proves that the proposed model could learn the relationship between the final angle and the required quantum state transformations.

\begin{figure}[t]
\centering
\includegraphics[width=0.95\columnwidth]{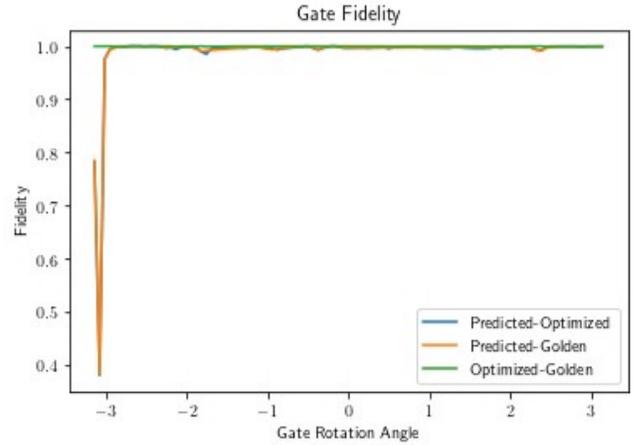}
\caption{Gate fidelity of forward model. Gates from ML model (``Predicted''), Juqbox optimizer (``Optimized''), and golden mathematically-derived (``Golden'') gates.}
\label{fig:fidelity}
\end{figure}

We adopted the gate fidelity in Eq.~\ref{eq:QuantumFidelity} as our last tool to quantify the quality of our results further. Figure~\ref{fig:fidelity} shows the fidelity when comparing the gates produced by our model, the traditional software optimizer, and mathematically-derived ground truth. Again, it is evident that the ML model performs poorly around $\beta{}=-\pi$, suggesting that the $-\pi$ row of the training dataset be removed so that the sign of $\beta$ values is consistent with the direction of the resultant rotation transformation.

\begin{figure}[t]
\centering
\includegraphics[width=0.95\columnwidth]{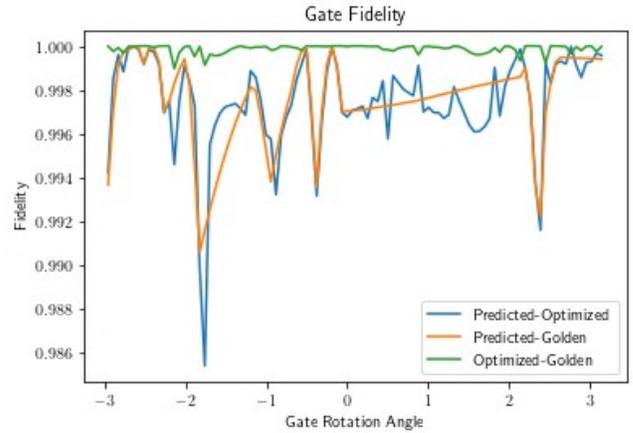}
\caption{Gate fidelity of forward model. Outlier results for smallest three $\beta$ removed.}
\label{fig:fidelity2}
\end{figure}

Figure~\ref{fig:fidelity2} shows the same results with the first three $\beta{}$ removed. The traditional software optimizer was configured to ensure fidelity greater than 0.999 between its results and the ground truth, as visible in the figure. It is interesting to note how the curve indicating overlap between the ML model outputs and the ground truth appears as a ``smoothed'' version of the overlap between the ML model outputs and the optimizer. Such a result suggests that the ML model can ignore lots of the random noise from the optimizer while effectively learning how to produce the target gate. The ML model produces gates with a gate fidelity of over $0.99$ for all remaining $\beta$, suggesting practical usefulness.

\section{Model codesign with hls4ml}

hls4ml is an open-source framework for the codesign of optimized neural networks and deployment on field-programmable gate arrays (FPGAs) and custom hardware (ASIC)~\cite{hls4ml}. At its core, hls4ml translates machine-learning models from common open-source software frameworks such as Tensorflow into a register-transfer level (RTL) implementation using high-level synthesis (HLS) tools~\cite{nane2015survey}.

The hls4ml framework originates from the Fast Machine Learning for Science community~\cite{deiana2022applications}, whose focus is the development of tools for scientific applications. The framework includes tools for the design-space exploration and the final FPGA or ASIC implementations. The resulting hardware implementations are configurable, spatial dataflow architectures tailored for speed and efficiency with extreme flexibility in the data-type precision~\cite{hendrik2022opensource}.

In hls4ml, a designer can trade off the performance (i.e., latency and throughput) and resource utilization for a model by varying the parallelization of the algorithm via several configuration parameters. For example, the \emph{reuse factor} (RF) parameter controls how many times each multiplier resource is used in the final hardware implementation: a designer with the goal of low latency will choose the lowest RF value.

\begin{figure}[t!]
\centering
\includegraphics[width=0.95\columnwidth]{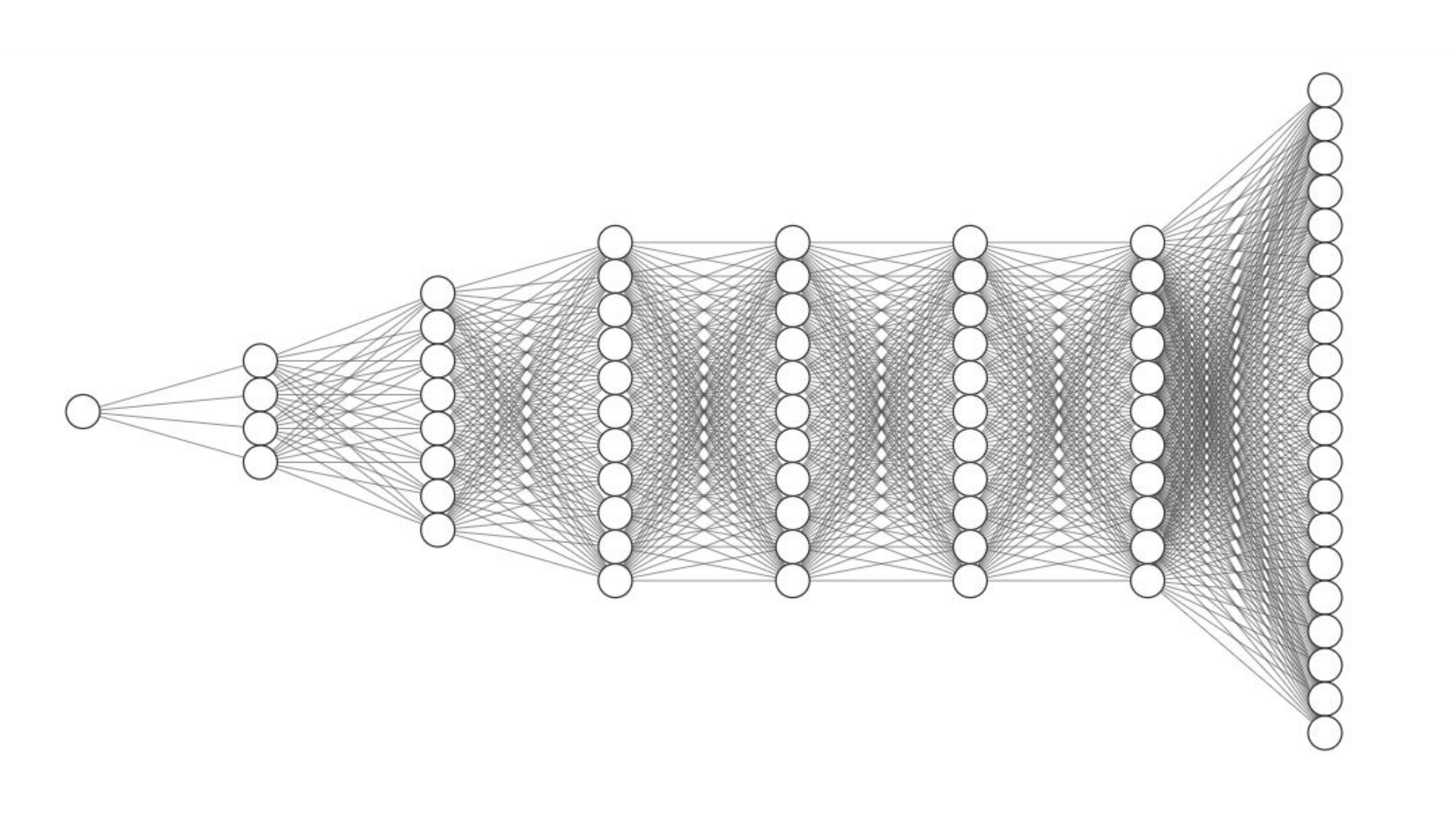}
\caption{A smaller multi-layer perceptron with six hidden layers and 783 parameters.\label{fig:qmodel1}}
\end{figure}

In this work, we used hls4ml, and additionally, we combined quantization-aware training in QKeras~\cite{qkeras} with model- and hardware-centric optimizations to find models that simultaneously accomplish the goals of high accuracy, low latency, and low FPGA-resource usage. We aimed to minimize resources such as onboard FPGA memory (BRAM), digital signal processing (arithmetic) blocks (DSPs), registers (flip-flops, or FFs), and programmable logic (lookup tables, or LUTs).

Our experiments targeted three off-the-shelf FPGA-based development boards with decreasing resources, thus making the codesign effort increasingly challenging: a Digilent Genesys 2, an Avnet Ultra96-V2, and a Digilent Arty A7-100T.

\subsection{Initial hardware implementation}
The Digilent Genesys 2 board is a development board equipped with an AMD/Xilinx Kintex-7 FPGA chip (\texttt{xc7k325tffg900-2}), which comes with 890 BRAMs, 840 DSPs, 407,600 FFs, and 203,800 LUTs. We passed as input of hls4ml the model developed in Sec.~\ref{sec:model_development} with a target clock period of 5 ns, a reuse factor of 1, and fixed-point precision with overall 16 bits for word width. The resulting implementation was a pipeline with a latency\footnote{The latency of a pipeline is the time required for one input to pass through the system from start to end.} of 35 cycles and initiation interval\footnote{The initiation interval of a pipeline is the time necessary for the system to process the next input (or to produce the next output).} of 1. The synthesized model used 54\% of the available DSPs and 13\% of available LUTs (other resources are not discussed as their utilization remains well below availability in this and all further experiments described herein).

\subsection{Shrinking the model}\label{sec:qmodel1}
The Avnet Ultra96-V2 is a smaller development board equipped with an AMD/Xilinx Zynq UltraScale MPSoC (\texttt{xczu3eg-sbva484-1-e}). This SoC combines four ARM Cortex A53 cores and programmable logic that offers 432 BRAMs, 360 DSPs, 141,1120 FFs, and 70,560 LUTs. The synthesis of our previous model returned a DSP utilization of 220\% and LUT utilization of 38\%. Thus, we took additional steps to reduce the resource requirements of our hardware implementation. We removed one hidden layer and some nodes from the remaining hidden layers. The resulting model shown in Fig.~\ref{fig:qmodel1} had a total of 783 parameters.

To reduce the final resource requirements on FPGA, besides reducing the overall number of model parameters, we could have increased the reuse factor in the hls4ml configuration, but we decided to keep the initial value of one. While a larger reuse factor would significantly decrease the size of the implementation, doing so also decreases parallelism, impacting latency and throughput.

Instead, to address the resource usage, we opted for a more aggressive quantization during the quantization-aware training with QKeras. We analyzed the model inputs and outputs, and for each layer, we chose a fixed-point precision with a 2-bit-integer part and either 10- or 11-bits for the overall word width. The new model trained in QKeras has an MSE of $6.6093\times{}10^{-8}$, which is slightly worse than the original model from Sec.\ref{sec:model_development}. But such a change is significant for the underlying FPGA-resource mapping since multipliers on 12 and fewer bits get mapped on LUTs rather than the scarcer DSPs. The resulting model fitted excellently within the Ultra96-V2 constraints: it uses 41\% of DSPs and 58\% of LUTs. 

\begin{figure}[t]
\centering
\includegraphics[width=0.95\columnwidth]{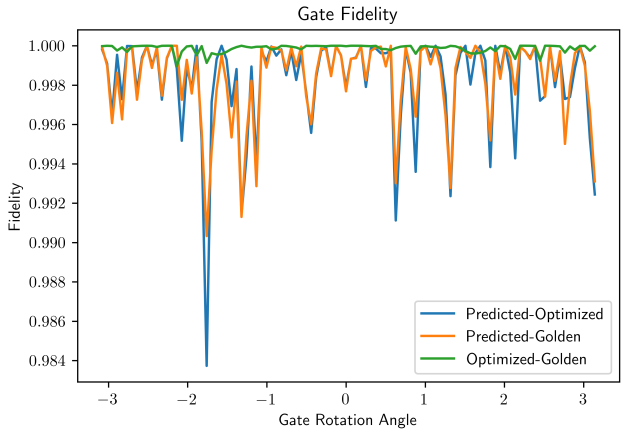}
\caption{Gate fidelity of the final implementation. Predicted-golden fidelity exceeds $0.99$ for all angles ($\beta{}$ values).}
\label{fig:fidelity_qmodel1_wa3}
\end{figure}

\subsection{Final implementation}
The third board we tested is an Arty A7-100T with an AMD/Xilinx Artix-7 chip (\texttt{xc7a100t-csg324-1}). This chip has fewer resources but shows similarities to a future eFPGA platform of interest: 270 BRAMs, 240 DSPs, 126,800 FFs, and 63,400 LUTs.

We adopted the same MLP architecture as in Fig.~\ref{fig:qmodel1}. However, we chose a mixed approach to balance DSP and LUT usage and the overall model accuracy. We used word lengths of 14, 12, 12, and 16 bits (and thus DSPs) for the first three layers and the last layer, respectively, while the word length for the remaining layers was 10 (and thus implemented in LUTs). The number of bits for the integer part was two for all layers, except the output layer, which has zero bits for the integer part.
The MSE measured on the test set was $5.7473\times{}10^{-8}$.

To fit the board-resource constraints, after the quantization-aware training phase, we manually edited the hls4ml-generated design. We set a smaller 10- and 12-bit-word length for the third and output layer, respectively. Such changes decreased the resource requirements while largely preserving performance.
The synthesized implementation fit the Artix-7 chip with a 99\% DSP utilization and 62\% LUT utilization. The resulting implementation was a pipeline with an initiation interval of 1 and a latency of 35 cycles. Looking at Figure~\ref{fig:fidelity_qmodel1_wa3}, we see that the fidelity between the gates produced by the model and the target gates is above $0.99$ for all $\beta$.

\section{Further Considerations}

In this section, we discuss some additional topics and future lines of research.

\subsection{Even higher fidelity}
Our final model achieves a gate fidelity of over $0.99$ for all $\beta$ examined. However, it is always desirable to increase the fidelity even further. For example, our training data reaches fidelity of over $0.999$ for all $\beta{}$, so the question arises: Would it be possible for a model trained on this data to achieve the same level of performance? Various solutions exist that can be combined with our current codesign flow based on hls4ml.
We can employ AutoML solutions to automatically discover a more performant model. These approaches are also known as neural architecture search (NAS)~\cite{elsken2019neural}. Tools like AutoKeras~\cite{autokeras} use a process of searching through neural network architectures to best address the optimal modeling task.

\begin{figure*}[t!]
\centering
\includegraphics[width=0.95\textwidth]{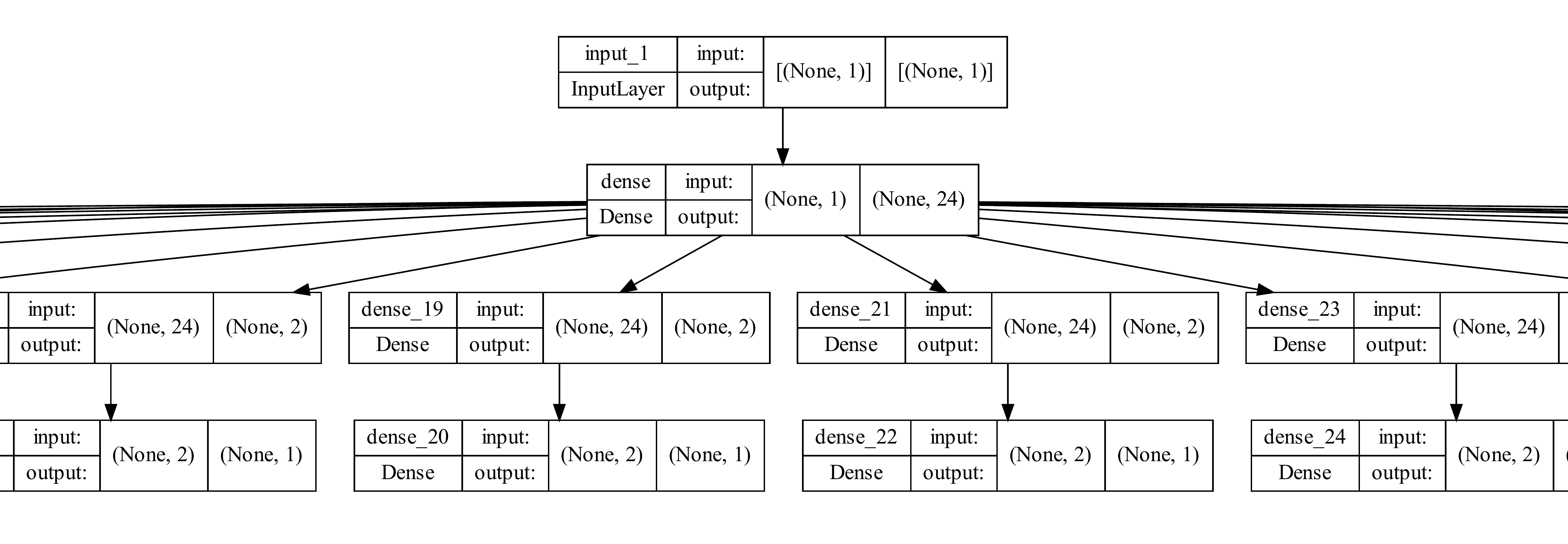}
\caption{Branched model design. A central crop of the architecture diagram is shown for clarity (the full diagram is available within the documentation repo). The figure continues to the left and right in a similar pattern for all 20 pulse parameters ($\alpha$).}
\label{fig:branchedmodel}
\end{figure*}

\begin{figure}[t]
\centering
\includegraphics[width=0.95\columnwidth]{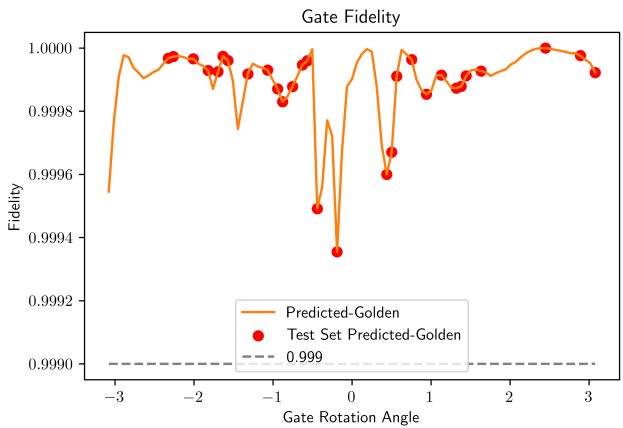}
\caption{Predicted-golden gate fidelity of large model. Test set points are marked by red circles, showing the ability of the model to interpolate. Fidelity exceeds $0.999$ for all test set $\beta{}$.}
\label{fig:fidelity_largemodel_testset_scatter}
\end{figure}

We have also observed that perhaps the function to compute each $\alpha$ differs greatly across the pulse parameters. Thus, we can devise a branched model in which each output is preceded by a hidden layer of its own, with a shared hidden layer at the top connected to the input. We started investigating such a model as in Fig.~\ref{fig:branchedmodel}, but we are still in the early phases of the analysis. 
Similarly, we are investigating larger models (up to 20 times the current number of parameters) that would require larger FPGAs but provide one order for magnitude lower MSE and fidelity that exceeds 0.999, as shown in Fig.~\ref{fig:fidelity_largemodel_testset_scatter}.

\subsection{Implementations with lookup tables}\label{sec:lut}

While it is significant that a machine-learning model can achieve comparable accuracy to a specific dataset, a designer may wonder why we cannot simply store the training dataset in a lookup table and use as an index the $\beta$ value so we can obtain the optimized $\alpha$ values when needed. Admittedly, such an implementation would take up a smaller area and have lower latency for this particular gate.

However, the desired parameter domain may be much larger for more complicated gates. For example, in this initial set of experiments, we have only one gate parameter $\beta{}$, which ranges from $-\pi$ to $\pi$. However, consider another gate that performs a rotation to \textit{any} point on the Bloch sphere. The expected number of entries in the lookup table will increase exponentially as a function of the number of input parameters if the same per-parameter granularity is maintained. With an implementation based on neural networks, however, as we are still outputting only 20 pulse parameters, we may expect the network resource requirements to grow slower. For example, one could even imagine that we duplicate the network to produce 20 pulse parameters independently for each input parameter. Then, the designer can use another network to combine these pulse parameters into one final set of pulse parameters. Such a network architecture grows linearly in the number of gate parameters.
Furthermore, these networks allow for interpolation and perhaps extrapolation. While a lookup table is constrained to the values stored, the ``intelligence'' of a neural network allows it to reasonably interpolate the pulse parameters for input gate parameters not seen during training.

Current quantum computing systems implemented using lookup tables will round gate parameters to the closest matching entry in the table \cite{liang20221}. The result is the application of lower-fidelity gates for the desired transformations, as the gate employed will have been optimized for different gate parameters than those of the desired gate. As a result, the accuracy of algorithmic results will decrease. For example, with the quantum approximate optimization algorithm (QAOA), we would expect the optimization results to be of lower quality, as gates are poorly approximated at each step~\cite{qaoa}.

One could also implement linear interpolations to guess the unknown pulse parameters starting from the pulse parameters of a known rotation angle. Consider optimizing $\alpha$ at the rotation angle $-\pi$ and using that $\alpha$ as initial guess for the next rotation angle. Then, to obtain $\alpha$ for intermediate rotation angles, one may use linear interpolation. We leave the comparison of accuracy of this method to the proposed ML technique for future investigation.

\section{Conclusions}

To summarize the main contributions of this work, we have developed a neural network capable of predicting pulse parameters for a quantum gate (specifically, a rotation on the Bloch sphere around the x-axis) to the fidelity of $0.99$ and shown an FPGA implementation of this network that fits a small off-the-shelf development board. This implementation exhibits a latency of 175 ns and is pipelined with an initiation interval of 1 and a clock period of 5 ns. Our numerical and experimental setup is publicly available at \cite{ml4quantum}. 

Preliminary results have shown that defining a machine learning model with a fidelity higher than $0.999$ is possible. We leave the FPGA deployment of such a model for future investigation together with the cost analysis of extending the current workflow, scaling analysis of the method for more complicated devices, the training with further-optimized training data, the augmentation of the dataset with data generated using additional random seeds, the design of a custom eFPGA-based system, the performance evaluation in extreme low-temperature environments, the comparison with lookup-table-based solutions, and the implementation of more complicated gates.

One may imagine that such models could be extended to control multi-qubit systems performing complex tasks. For example, application-specific hardware accelerators could be designed to capture multi-stage domain-specific operations in gates generated on-the-fly. Such abilities would allow for cheaper, more precise, and faster quantum computing systems that will be used to solve the problems of tomorrow. By moving the computation into the colder domains of the quantum computing system, the number of wires leading to the warmer domains can be reduced significantly (as we no longer need wires to transmit $\alpha$ values calculated externally). Such wires are costly in the amount of energy they dissipate within the cold domains which are managed by machines of limited cooling capacity~\cite{xue2021cmos}.

\subsection{Acknowledgements}

We thank Matt Otten for discussions on pulse parameters and neural networks. This work was conducted as part of the Spring 2022 seminar course ``CSEE E6868 - Embedded Scalable Platforms'' at Columbia University. DX thanks the TA Joseph Zuckerman for his help throughout the semester. AB\"O and GNP were partially supported by the DOE/HEP QuantISED program grant ``HEP Machine Learning and Optimization Go Quantum,'' identification number 0000240323 and the DOE/HEP QuantISED program grant ``QCCFP-QMLQCF Consortium,'' identification number DE-SC0019219. This manuscript has been authored by Fermi Research Alliance, LLC under Contract No. DE-AC02-07CH11359 with the U.S. Department of Energy, Office of Science, Office of High Energy Physics. We acknowledge the Fast Machine Learning collective as an open community of multi-domain experts and collaborators. This community was important for the development of this project.

\bibliographystyle{IEEEtran}
\bibliography{ref}

\end{document}